\documentclass{elsarticle}
\usepackage[utf8]{inputenc}
\usepackage{latexsym,amsthm,amsmath,amssymb}
\usepackage[]{paralist}
\usepackage{tikz}
\usetikzlibrary{arrows, arrows.meta}
\usetikzlibrary{shapes.arrows,patterns,decorations.pathreplacing}
\usetikzlibrary{calc}

\usepackage{hyperref}
\usepackage{tabularx}
\usepackage[capitalise]{cleveref}
\usepackage{microtype}

\usepackage{todonotes}

\newcommand{\overbar}[1]{\mkern 3.5mu\overline{\mkern-3.5mu#1\mkern-1.5mu}\mkern 1.5mu}

\title{Graphs without a partition into two proportionally dense subgraphs}
\date{}

\author[paris]{Cristina Bazgan}
\ead{bazgan@lamsade.dauphine.fr}
\author[portsmouth]{Janka Chleb\'\i kov\'a}
\ead{janka.chlebikova@port.ac.uk}
\author[portsmouth]{Cl\'ement Dallard}
\ead{clement.dallard@port.ac.uk}

\address[paris]{Universit\'e Paris-Dauphine, PSL Research University, CNRS,  LAMSADE, Paris, France}
\address[portsmouth]{School of Computing, University of Portsmouth, Portsmouth, United Kingdom}

\hypersetup{
  bookmarksopen=false,
  pdftitle={Graphs without a partition into two proportionally dense subgraphs},
  pdfauthor={},
  pdftoolbar=false, 
  pdfmenubar=true, 
  pdfcreator=\LaTeX,
  pdfpagelayout=SinglePage, 
  pdffitwindow=true, 
  pdfkeywords={graphs,community},
  colorlinks=true,
  linkcolor={red!60!black},
  citecolor={blue!50!black},
  urlcolor={blue!80!black}
}

\newtheorem{theorem}{Theorem}

\theoremstyle{definition}
\newtheorem{definition}{Definition}

\newcommand{\NP}{\textsc{NP}}

\begin{document}
\begin{frontmatter}

\begin{abstract}
A \emph{proportionally dense subgraph} (PDS) is an induced subgraph of a graph such that each vertex in the PDS is adjacent to proportionally as many vertices in the subgraph as in the rest of the graph.
In this paper, we study a partition of a graph into two proportionally dense subgraphs, namely a \emph{$2$-PDS partition}, with and without additional constraint of connectivity of the subgraphs.
We present two infinite classes of graphs: one with graphs without a $2$-PDS partition, and another with graphs that only admit a disconnected $2$-PDS partition.
These results answer some questions proposed by Bazgan et al. [Algorithmica 80(6) (2018), 1890--1908].
\end{abstract}
\end{frontmatter}

\section{Introduction}

The problems of partitioning a graph into two parts have been intensively studied with various objective functions and constraints.
Let's mention at least two such \NP-hard problems.
The \textsc{Satisfactory Partition} problem \cite{bazgan2006satisfactory} asks whether a graph can be partitioned into two parts such that every vertex is adjacent to more vertices in its own part than in the other.
In the \textsc{Maximally Balanced Connected Partition} problem, the task is to partition a graph into two connected subgraphs such that the size of the smallest  subgraph is maximised \cite{chlebikova1996approximating}.

The notion of \emph{proportionally dense subgraph} is closely related to the notion of \emph{community} as introduced in \cite{Ols13}.
\citeauthor{Ols13} defines a \emph{community structure} as a partition of the vertices into communities, where a part, \textit{i.e.}\ an induced subgraph (with at least $2$ vertices), is a \emph{community} if and only if each vertex has proportionally as many neighbours in its community than in any other community.
In \cite{BCP17}, the authors investigate the notion of $2$-community structure as a community structure with exactly two parts.
We use the same definition (up to the special case where a community is of size one) to define a $2$-PDS partition.

So far, only few results are known about the existence of a $2$-PDS partition in a graph, and the complexity of finding one.
It has been proved in \cite{ECP13} that deciding if a graph contains a $2$-PDS partition with both PDS's of the same size is \NP-complete.
On trees \cite{BCP17,ECP13} and graphs with maximum degree $3$ or minimum degree $n-3$, ($n$ the order of the graph) a connected $2$-PDS partition always exists and can be found in polynomial time \cite{BCP17}.
The results extensively use the connectivity of the PDS's.
To find a connected $2$-PDS partition in a tree, one can prove that there exists an edge such that its removal yields two connected PDS's.
If a graph has a maximum degree at most $3$, a greedy algorithm keeps decreasing the size of a cut under some constraints and the removal of the final cut describes two connected PDS's.

Another problem related to the notion of PDS is the \textsc{Max PDS} problem.
In this problem, the goal is to determine the size of a maximum PDS (with regard to the number of vertices) in a given graph.
Hence, only the vertices inside the PDS must be satisfied.
In \cite{maxpds}, the authors prove that \textsc{Max PDS} is \NP-hard on bipartite and split graphs, and propose a  polynomial-time $(2-\frac{2}{\Delta+1})$-approximation algorithm, where $\Delta$ is the maximum degree of the graph.
They also show that deciding if a subset of vertices can be a (proper) subset of the vertices of a PDS is \textsc{co-NP}-complete on bipartite graphs.

\paragraph{Our contributions}
In \cref{section: PDS and PDS partitions}, we formally define the concepts of proportionally dense subgraphs and PDS partitions, and outline the known results about the \textsc{$2$-PDS partition} problem.
Then, we construct an infinite family of graphs without a $2$-PDS partition in \cref{subsection: no 2-PDS partition}.
As far as we know, these are the first negative results regarding the existence of a $2$-PDS partition.
We also give examples of graphs without a $2$-PDS partition that do not belong to the family.
In \cref{subsection: disconnected 2-PDS partition} we present another infinite family of graphs without a connected $2$-PDS, but with a disconnected one.

\section{Proportionally dense subgraphs}\label{section: PDS and PDS partitions}

All graphs in this paper are simple.
Given a graph $G=(V,E)$ and a subset of vertices $S \subset V$, $\overbar{S}$ refers to the set $V \setminus S$.
For a vertex $u\in V$, $N(u)$ represents the set of neighbours of $u$, $d(u) := |N(u)|$ is the degree of $u$, and $d_S(u) := |N(u) \cap S|$ denotes the degree of $u$ in $S$.
We say that a vertex $u \in V$ is \emph{universal} if it is connected to all other vertices of the graph, that is, $d(u) = |V|-1$.

The density of a subgraph on a vertex set $S \subseteq V$ is usually defined as $\frac{|E(S)|}{|S|}$, where $E(S)$ is the set of edges in the subgraph.
The problem of finding a subgraph of maximum density can be solved in polynomial time \cite{Goldberg:CSD-84-171}, but it becomes \NP-hard when at least, or exactly, $k$ vertices must belong to the subgraph \cite{feige2001dense,asahiro2002complexity,KhullerSaha2009}.

In this paper, we introduce the notion of \emph{proportionally dense subgraph} (PDS), which captures both the size of the subset and the number of neighbours.

\begin{definition}\label{definition: PDS}
For a graph $G=(V,E)$, a \emph{proportionally dense subgraph} of $G$ is an induced subgraph on a vertex set $S \subset V$ such that each vertex $u \in S$ is \emph{satisfied} in $S$, that is,
\begin{equation*}
      |\overbar{S}|\cdot d_{S}(u) \geq (|S|-1)\cdot d_{\overbar{S}}(u)\,,
           \text{~or, equivalently,~}
      (|V|-1)\cdot d_{S}(u) \geq (|S|-1)\cdot d(u)\,.
\end{equation*}
 Note that if $|S| \geq 2$, then we can rewrite the inequalities as
\begin{equation*}
		\frac{d_S(u)}{\mathstrut{|S|-1}} \geq \frac{d_{\overbar{S}}(u)}{|\mathstrut{\overbar{S}|}}
		\text{~or, equivalently,~}
        \frac{d_{S}(u)}{|S|-1} \geq \frac{d(u)}{|V|-1}\,.
\end{equation*}
The proof of the equivalence can be found in \cite{BCP17}. Note that a subgraph containing a single vertex is also a PDS, but obviously a PDS cannot be the entire graph.

\end{definition}
\usetikzlibrary{shapes.arrows,patterns,decorations.pathreplacing}

\begin{definition}\label{definition:PDS partition}
	A \emph{$2$-PDS partition} of a graph $G=(V,E)$ is a partition $\Pi=\{S_1, S_2\}$ of $V$ such that $S_1$ and $S_2$ induce two PDS's in $G$.
\end{definition}

In this paper, we address the problem of deciding if a graph admits a $2$-PDS partition.
Notice that a PDS doesn't necessarily need to be connected.
Therefore we also consider the problem of deciding if a graph has a \emph{connected $2$-PDS partition}, that is, a $2$-PDS partition whose PDS's are connected subgraphs.

If a graph is disconnected, both problems become trivial, hence we assume that all graphs are connected.

\section{Infinite classes of graphs}

\subsection{Graphs without \texorpdfstring{$2$}{2}-PDS partition}\label{subsection: no 2-PDS partition}
The question about the existence of graphs without a $2$-PDS was left open in \cite{BCP17}.
To the best of our knowledge, no graphs without a $2$-PDS partition were known.
In this section we present an infinite class $\mathcal{G}$ (see \cref{construction:bad family}) of graphs with even number of vertices without a $2$-PDS partition.

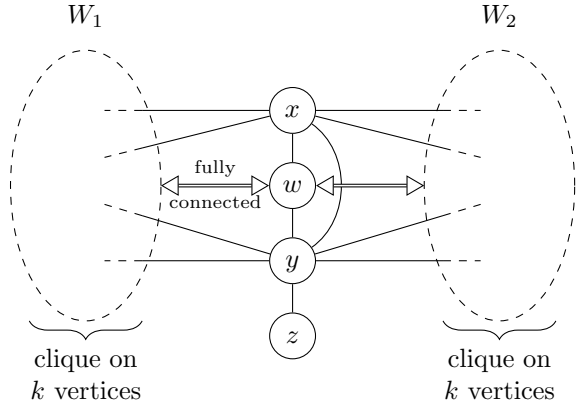
\begin{figure}[htbp]
\centering
\usetikzlibrary{arrows, arrows.meta}
	\begin{tikzpicture}[scale=1] 
	\begin{scope}
		\tikzset{every node/.style={circle,
				inner sep=0pt,
				text width=5mm,
				align=center,
				draw=black,
				transform shape,
				fill=white}}

		\node (z) at (0,0) {$\mathstrut z$};
		\node (y) at (0,1) {$\mathstrut y$};
		\draw (y) to (z);

		\node (x) at (0,3) {$\mathstrut x$};
		\node (w) at (0,2) {$\mathstrut w$};
		\draw (y) to (w);
		\draw (w) to (x);
		\draw (y) to [bend right=55] (x);
	\end{scope}
	\draw[dashed] (-2.75,2) ellipse (1cm and 1.8cm);
	\node (w1) at (-2.75,4.25) {$\mathstrut W_1$};
	\draw [decorate,decoration={brace,amplitude=7pt}] (-2,0.2) -- (-3.50,0.2);
	\draw node[draw=none,rectangle, text width=2cm, text centered] at (-2.75,-0.5) {clique on $k$ vertices};
	\draw[dashed] (2.75,2) ellipse (1cm and 1.8cm);
	\node (w1) at (2.75,4.25) {$\mathstrut W_2$};
    \draw [decorate,decoration={brace,amplitude=7pt}] (3.50,0.2) -- (2,0.2);
	\draw node[draw=none,rectangle, text width=2cm, text centered] at (2.75,-0.5) {clique on $k$ vertices};

	\draw[] (x) to (-2,3);
	\draw[dashed] (-2,3) to ++(-.5,0);
	\draw[] (x) to (-2,2.5);
	\draw[dashed] (-2,2.5) to ++(-.5,-.125);

	\draw[] (y) to (-2,1);
	\draw[dashed] (-2,1) to ++(-.5,0);
	\draw[] (y) to (-2,1.6);
	\draw[dashed] (-2,1.6) to ++(-.5,0.15);


	\draw[] (x) to (2,3);
	\draw[dashed] (2,3) to ++(.5,0);
	\draw[] (x) to (2,2.5);
	\draw[dashed] (2,2.5) to ++(.5,-.125);

	\draw[] (y) to (2,1);
	\draw[dashed] (2,1) to ++(.5,0);
	\draw[] (y) to (2,1.6);
	\draw[dashed] (2,1.6) to ++(.5,0.15);

	\draw[{Triangle[open,scale=1]}-{Triangle[open,scale=1]},double] (w) --
	node[midway,above, font = \scriptsize] {fully}  node[midway,below,font = \scriptsize] {connected}
	(-1.75,2);
	\draw[{Triangle[open,scale=1]}-{Triangle[open,scale=1]},double] (w) --
(1.75,2);


%

	\end{tikzpicture}
\caption{A schematic representation of a graph in $\mathcal{G}$.}\label[figure]{fig:antisocial graph}
\end{figure}

\begin{definition}\label{construction:bad family}
	Let $\mathcal{G}$ be the class of graphs such that, if $G=(V,E) \in \mathcal{G}$, then
	\begin{compactitem}
		\item $V=W_1 \cup W_2\cup \{w, x, y, z\}$, where $W_1$, $W_2$ are cliques of the same size $k$, $k\geq 3$, and $\{w, x, y\}$ is a clique of size $3$;
		\item $w$ is adjacent to all vertices in $W_1\cup W_2$, and $z$ is only adjacent to $y$\;
		\item $1 \leq d_{W_1}(x) = d_{W_2}(x) \leq k-1$ and $2 \leq d_{W_1}(y) = d_{W_2}(y) \leq k-1$;
		\item $|W_i \cap \left( N(x) \cup N(y) \right)| > \frac{3k}{k+3}$ for each $i \in \{1,2\}$;
		\item there exist vertices $\alpha, \beta \in W_1 \cup W_2$ such that $\alpha \in  N(y)\setminus N(x) $, and $\beta \in N(x) \cap N(y)$;
		\item there is no edge between the vertex sets  $W_1$ and $W_2$.
	\end{compactitem}
\end{definition}

\begin{figure}[htbp]
\centering
    \begin{tikzpicture}[scale=1]           
    \tikzset{every node/.style={circle,
				inner sep=0pt,
				text width=5mm,
				align=center,
				draw=black,
				transform shape,
				fill=white}}
	
	\node (z) at (0,0) {$\mathstrut z$};
	\node (y) at (0,1) {$\mathstrut y$};
	\draw (y) to (z);
	
	\node (x) at (0,3) {$\mathstrut x$};
	\node (w) at (0,2) {$\mathstrut w$};
	\draw (y) to (w);
	\draw (w) to (x);
	\draw (y) to [bend right=45] (x);
	
	\node (a) at (-2.25,1) {$\mathstrut a$};
	\node (b) at (-1.50,2) {$\mathstrut b$};
	\node (c) at (-2.25,3) {$\mathstrut c$};
	\draw (a) to (b);
	\draw (b) to (c);
	\draw (c) to (a);
	
	\node (a') at (2.25,1) {$\phantom{'}\mathstrut a'$};
	\node (b') at (1.5,2) {$\phantom{'}\mathstrut b'$};
	\node (c') at (2.25,3) {$\phantom{'}\mathstrut c'$};
	\draw (a') to (b');
	\draw (b') to (c');
	\draw (c') to (a');
	
	\foreach \u in {a,b,c,a',b',c'} {
		\draw (w) to (\u);	
	}
	
	
	\foreach \u in {a,a',b,b'} {
		\draw (y) to (\u);
	}
	
	\foreach \u in {b,b'} {
		\draw (x) to (\u);
	}
\end{tikzpicture}%
    \hspace{0.125cm}%
    \begin{tikzpicture}[scale=1]

    \node[draw=none,fill=none,minimum width=0mm,inner sep=0pt,align=none,outer sep=0mm,text width=0pt] (toa) at (-2.85,2.85) {};
    \node[draw=none,fill=none,minimum width=0mm,inner sep=0pt,align=none,outer sep=0mm,text width=0pt] (toa') at (2.85,2.85) {};

    \tikzset{every node/.style={circle,
				inner sep=0pt,
				text width=5mm,
				align=center,
				draw=black,
				transform shape,
				fill=white}}

	\node (z) at (0,0) {$\mathstrut z$};
	\node (y) at (0,1) {$\mathstrut y$};
	\draw (y) to (z);
	
	\node (x) at (0,2) {$\mathstrut x$};
	\node (w) at (0,3.25) {$\mathstrut w$};
	\draw (y) to (x);
	
	\node (a) at (-2.25,1) {$\mathstrut a$};
	\node (b) at (-1.5,2) {$\mathstrut b$};
	\node (c) at (-2.25,3) {$\mathstrut c$};
	\draw (a) to (b);
	\draw (b) to (c);
	\draw (c) to (a);
	
	\node (a') at (2.25,1) {$\phantom{'}\mathstrut a'$};
	\node (b') at (1.5,2) {$\phantom{'}\mathstrut b'$};
	\node (c') at (2.25,3) {$\phantom{'}\mathstrut c'$};
	\draw (a') to (b');
	\draw (b') to (c');
	\draw (c') to (a');
	

    \draw (w) to [out=10, in=105] (toa');
    \draw (toa') to [out=180+105, in=65] (a');
    \draw (w) to [out=180-10, in=180-105] (toa);
    \draw (toa) to [out=-105, in=180-65] (a);
	\draw (w) to (b);
	\draw (w) to (b');
	\draw (w) to (c);
	\draw (w) to (c');
	\draw (w) to (x);
  	\draw (w) to [out=20,in=130] (3.1,3.1) to [out=180+130, in=50] (2.8,0.8) to [out=50+180, in=-20] (y);

	
	\foreach \u in {a,a',b,b'} {
		\draw (y) to (\u);
	}
	
	\foreach \u in {b,b'} {
		\draw (x) to (\u);
	}
\end{tikzpicture}
\caption{A planar graph from $\mathcal{G}$ with 10 vertices without a $2$-PDS partition.
On the left, its schematic representation as in \cref{fig:antisocial graph}; on the right,  its planar representation.}\label[figure]{fig:planar}
\end{figure}
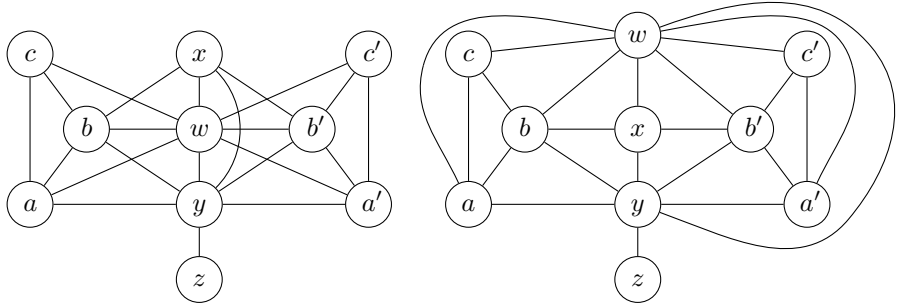

Note that the smallest graphs in $\mathcal{G}$ have $10$ vertices, and one of them is planar (see \cref{fig:planar}).

\begin{theorem}\label{main}
	All graphs in $\mathcal G$ do not have a $2$-PDS partition. 
\end{theorem}
\begin{proof}
	\setlength{\pltopsep}{3pt}
	\setlength{\plitemsep}{3pt}
	\setdefaultleftmargin{1.7em}{1em}{1.7em}{1.7em}{1em}{1em}
Let $G=(V,E)$ be a graph in $\mathcal G$.
Firstly, notice that there is no $2$-PDS partition $\{A,B\}$ in $G$ such that $|A| = 1$ or $|B| = 1$.
Without loss of generality, suppose by contradiction that $A=\{v\}$ for some vertex $v \in V$, and notice that the neighbour of $v$ in $B$ must be a universal vertex in order to be satisfied.
Since $G$ does not contain a universal vertex, there is no $2$-PDS partition $\{A,B\}$ in $G$ with $|A|=1$ or $|B|=1$.
Hence, assume that $|A|, |B| \geq 2$.

Observe that the vertex $z$ is satisfied if and only if it belongs to the same PDS as the vertex $y$. Hence, without loss of generality, we assume that $y,z \in B$.
In addition, the vertex $w$ has degree $|V|-2$ and is not connected to $z \in B$.
Hence, necessarily $w \in A$.

Now we prove that for any partition $\{A,B\}$ of $V$, where $w \in A$ and $y,z \in B$, there is at least one vertex which is not satisfied, hence there is no $2$-PDS partition in $G$.
For any partition $\{A,B\}$ of $V$, we denote by $A_i$ and $B_i$ the sets $A \cap W_i$ and $B \cap W_i$, respectively, for $i \in \{1,2\}$.
We split the proof into two cases: In the first case, we suppose that $B_1$ or $B_2$ is empty; in the second case, we assume that $B_1$ and $B_2$ are not empty.

\bigskip

\textbf{\underline{Case 1: $B_1=\emptyset$ or $B_2=\emptyset$}}

\medskip

Suppose first that $B_1 = \emptyset$ and $B \subseteq \{x,y,z\} \cup W_2$.
	\begin{compactitem}
		\item If $B_2 = \emptyset$, we have two possibilities:
		\begin{compactitem}
			\item if $x \in B$, then $B = \{x,y,z\}$ and $\beta \in A$ is not satisfied since $\frac{d_A(\beta)}{|A|-1}= \frac{k}{2k} < \frac{2}{3} = \frac{d_B(\beta)}{|B|}$;

			\item if $x \in A$, then $B = \{y,z\}$ and $\alpha \in A$ is not satisfied since $\frac{d_A(\alpha)}{|A|-1} = \frac{k}{2k+1} < \frac{1}{2} = \frac{d_B(\alpha)}{|B|}$.
		\end{compactitem}

		\item If $B_2 \neq \emptyset$ and $B_2 \neq W_2$,
		\begin{compactitem}
			\item Case $x \in B$.
			\begin{compactitem}
				\item If there exists $u \in A_2$ such that $u \in N(x) \cup N(y)$ and $u$ is satisfied, then we have:
				\begin{flalign*}
				\frac{|A_2|}{k+|A_2|} = \frac{d_A(u)}{|A|-1} \geq \frac{d(u)}{|V|-1} \geq \frac{k+1}{2k+3}\,,
				\end{flalign*}
				which implies that $|A_2| \cdot (k+2) \geq k \cdot (k+1)$, hence that $|A_2| > k-1$.
				A contradiction since $|A_2| \leq k-1$.

				\item Otherwise, for all $u \in A_2$, $u \notin N(x) \cup N(y)$. Hence, for any $u\in A_2$, if $u$ is satisfied then:
				\begin{flalign*}
				\frac{|A_2|}{k+|A_2|}=\frac{d_A(u)}{|A|-1}\geq \frac{d(u)}{|V|-1}=\frac{k}{2k+3}\,,
				\end{flalign*}
				which implies that $|A_2| \cdot (k+3) \geq k^2$, hence that $|A_2| \geq \frac{k^2}{k+3}$.
				Due to our assumptions about the graph, $|W_2 \cap \left( N(x) \cup N(y) \right)| > \frac{3k}{k+3}$.
				Thus, $k - \frac{3k}{k+3} > |W_2\setminus (N(x)\cup N(y))|\geq |A_2| \geq \frac{k^2}{k+3}$ which implies $k > k$, a contradiction.
			\end{compactitem}

			\item Case $x \in A$. Let $u\in A_2$.
			\begin{itemize}
				\item If $u \in N(y)\cap N(x)$ and $u$ is satisfied, then we have:
				\begin{flalign*}
				\frac{|A_2|+1}{k+|A_2|+1} = \frac{d_A(u)}{|A|-1}\geq \frac{d(u)}{|V|-1}=\frac{k+2}{2k+3}\,,
				\end{flalign*}
				which implies that $|A_2| \geq k-\frac{1}{k+1}$, and then $|A_2|\geq k$, a contradiction since $B_2\neq \emptyset$.
				\item If $u \in N(y)\setminus N(x)$, then $d_A(u)=|A_2|$ and $d(u)=k+1$.
				Therefore, similarly to the previous case, we obtain that $|A_2|\geq k + \frac{1}{k+2}$ and so $|A_2|> k$, a contradiction.

				\item If $u \in N(x)\setminus N(y)$, then:
				\begin{flalign*}
				\frac{|A_2|+1}{k+|A_2|+1} = \frac{d_A(u)}{|A|-1} \geq \frac{d(u)}{|V|-1} = \frac{k+1}{2k+3}\,,
				\end{flalign*}
				which implies that $|A_2| \cdot (k+2) \geq k^2-2$, hence $|A_2| \geq \frac{k^2-2}{k+2} > k-2$.
				Since assuming that there is a vertex in $A_2\cap N(y)$ leads to a contradiction (see previous cases), we can assume that $A_2\cap N(y)=\emptyset$. Then, since $d_{W_2}(y) \geq 2$, then $|W_2 \setminus N(y)| \leq k-2$.
				Thus $k-2 \geq |A_2| > k-2$, a contradiction.\\
				\item If $u \notin N(x)\cup N(y)$, then $d_A(u)=|A_2|$ and $d(u)=k+1$.
				Again, similarly to the previous case, we obtain $|A_2| > k $, a contradiction since $|B_2|\neq \emptyset$.
			\end{itemize}

		\end{compactitem}
		\item If $B_2=W_2$, then either $B = \{x, y, z\} \cup W_2$, and we have $|A|+2=|B|$ but $d_A(x) = d_B(x)$ thus $x$ is not satisfied, or $B = \{y, z\} \cup W_2$, and since $|A|=|B|$ we have: $\frac{d_B(y)}{|B|-1} < \frac{d_B(y)+1}{|B|} = \frac{d_A(y)}{|B|} =  \frac{d_A(y)}{|A|} $, thus $y$ is not satisfied.
	\end{compactitem}

We conclude that if there is a $2$-PDS partition in $G$, then $B_1 \neq \emptyset$.
The case $B_2 = \emptyset$ is similar, therefore if there is a $2$-PDS partition in $G$, then $B_2 \neq \emptyset$.

\bigskip

\textbf{\underline{Case 2: $B_1,B_2 \neq \emptyset$}}.

\medskip

	Without loss of generality, we suppose $|B_1| \leq |B_2|$. Let $u \in B_1$ and suppose that $u$ is satisfied in the partition $\{A,B\}$. We prove that in all cases, if $u$ is satisfied then it implies a contradiction with $|B_1| \leq |B_2|$.
\setdefaultitem{}{\textasteriskcentered}{}{}
\begin{compactitem}
\item If $x\in A$
		\begin{itemize}
		\item If $u\in N(x)\cap N(y)$ is satisfied, then:
         \begin{flalign*}
		\frac{|B_1|}{|B_1|+|B_2|+1} = \frac{d_B(u)}{|B|-1} \geq \frac{d(u)}{|V|-1} = \frac{k+2}{2k+3}\,,
		\end{flalign*}
		which implies that $|B_1| \cdot (k+1) \geq (|B_2|+1) \cdot (k+2)$, hence that $|B_1| > |B_2|$.
		A contradiction with the assumption that $|B_1| \leq |B_2|$, hence $u$ is not satisfied.
		
        \item If $u\in N(x)\setminus N(y)$, we have $d_B(u)=|B_1|-1$ and $d(u)=k+1$ and similarly we obtain $|B_1|\cdot (k+2)\geq |B_2|\cdot (k+1) +(3k+4)\geq |B_2|\cdot (k+1)+(|B_2|+4)> |B_2|\cdot (k+2)$, a contradiction since $|B_1| \leq |B_2|$.
        
        \item If $u\in N(y)\setminus N(x)$, we have $d_B(u)=|B_1|$ and $d(u)=k+1$ and similarly we obtain $|B_1|\cdot (k+2)\geq |B_2|\cdot (k+1) +(k+1)\geq |B_2|\cdot (k+1)+(|B_2|+1)> |B_2|\cdot (k+2)$, a contradiction since $|B_1| \leq |B_2|$.
		
		\item If $u\notin N(x)\cup N(y)$, we have $d_B(u)=|B_1|-1$ and $d(u)=k$ and similarly we obtain $|B_1|\cdot (k+3)\geq |B_2|\cdot k +3(k+1)\geq |B_2|\cdot k +3(|B_2|+1)> |B_2|\cdot (k+3)$, a contradiction since $|B_1| \leq |B_2|$.
		\end{itemize}
\item If $x\in B$
		\begin{itemize}
		\item If $u\in N(x)\cap N(y)$ is satsfied, then:
        \begin{flalign*}
		\frac{|B_1|+1}{|B_1|+|B_2|+2} = \frac{d_B(u)}{|B|-1} \geq \frac{d(u)}{|V|-1} = \frac{k+2}{2k+3}\,,
		\end{flalign*}
		which implies that $|B_1| \cdot (k+1) \geq |B_2| \cdot (k+2) + 1$, thus that $|B_1| > |B_2|$.
		A contradiction with the assumption that $|B_1| \leq |B_2|$, hence $u$ is not satisfied.
		
        \item If $u\in N(x)\setminus N(y)$ or $u\in N(y)\setminus N(x)$, we have $d_B(u)=|B_1|$ and $d(u)=k+1$ and similarly we obtain $|B_1| \cdot (k+2) \geq |B_2| \cdot (k+1) + 2(k+1)\geq |B_2| \cdot (k+1) + 2(|B_2|+1)>|B_2| \cdot (k+3)$, a contradiction since $|B_1|\leq |B_2|$.
        
		\item If $u\notin N(x)\cup N(y)$, we have $d_B(u)=|B_1|-1$ and $d(u)=k$ and similarly we obtain $|B_1|\cdot (k+3)\geq |B_2|\cdot k +4k+3\geq |B_2|\cdot \frac{k}{k+3} +4\cdot|B_2|+3>|B_2|\cdot (k+4)$, a contradiction since $|B_1|\leq |B_2|$.
		\end{itemize}
\end{compactitem}
\end{proof}

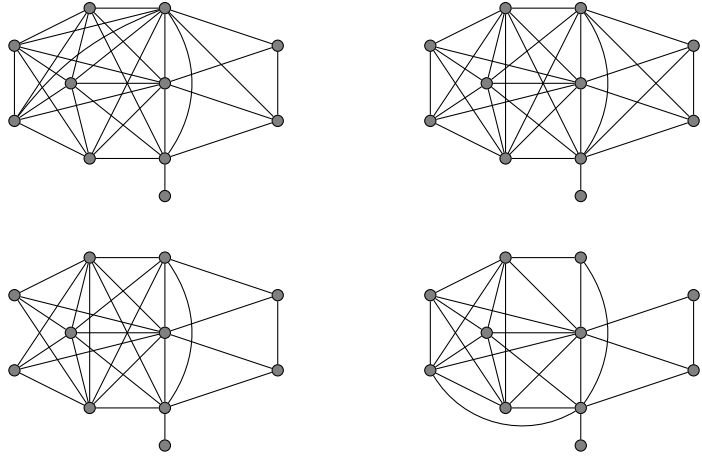
\begin{figure}[htbp]
    \centering
    \begin{tabular}{c@{\hskip 1.75cm}c}
    \begin{tikzpicture}[scale=1]
\tikzset{every node/.style={circle,
				inner sep=0pt,
				text width=1.5mm,
				align=center,
				draw=black,
				transform shape,
				fill=gray}}
		
    \node (z) at (0,0.5) {}; 
	\node (y) at (0,1) {}; 
	\draw (y) to (z);
	\node (x) at (0,3) {}; 
	\node (w) at (0,2) {}; 
	\draw (y) to (w);
	\draw (w) to (x);
	\draw (y) to [bend right=35] (x);
	
	\node (l1) at (-2,2.5) {};
	\node (l2) at (-2,1.5) {};
	\node (l3) at (-1,3) {};
	\node (l4) at (-1.25,2) {};
    \node (l5) at (-1,1) {};
    
    \foreach \u in {l3,l4,l5}{
        \draw (x) to (\u);
        \draw (y) to (\u);
    }
    \draw (x) to (l1);
    \draw (x) to [bend right=15] (l2);
    
    \draw (w) to [] (l1);
    \draw (w) to [] (l2);
    \draw (w) to [] (l3);
    \draw (w) to [] (l4);
    \draw (w) to [] (l5);
    
	\draw (l1) to (l2);
	\draw (l1) to (l3);
	\draw (l1) to (l4);
	\draw (l1) to (l5);
	\draw (l2) to (l3);
	\draw (l2) to (l4);
	\draw (l2) to (l5);
	\draw (l3) to (l4);
	\draw (l4) to (l5);
	
	\node (r1) at (1.5,2.5) {};
	\node (r2) at (1.5,1.5) {};
	
	\draw (r1) to (r2);
	\draw (w) to (r1);
	\draw (w) to (r2);
	\draw (x) to (r1);
	\draw (x) to (r2);
	\draw (y) to (r2);

\end{tikzpicture} & \begin{tikzpicture}[scale=1]
\tikzset{every node/.style={circle,
				inner sep=0pt,
				text width=1.5mm,
				align=center,
				draw=black,
				transform shape,
				fill=gray}}
		
    \node (z) at (0,0.5) {}; 
	\node (y) at (0,1) {}; 
	\draw (y) to (z);
	\node (x) at (0,3) {}; 
	\node (w) at (0,2) {}; 
	\draw (y) to (w);
	\draw (w) to (x);
	\draw (y) to [bend right=35] (x);
	
	\node (l1) at (-2,2.5) {};
	\node (l2) at (-2,1.5) {};
	\node (l3) at (-1,3) {};
	\node (l4) at (-1.25,2) {};
    \node (l5) at (-1,1) {};
    
    \foreach \u in {l3,l4,l5}{
        \draw (x) to (\u);
        \draw (y) to (\u);
    }

    \draw (w) to [] (l1);
    \draw (w) to [] (l2);
    \draw (w) to [] (l3);
    \draw (w) to [] (l4);
    \draw (w) to [] (l5);
    
	\draw (l1) to (l2);
	\draw (l1) to (l3);
	\draw (l1) to (l4);
	\draw (l1) to (l5);
	\draw (l2) to (l3);
	\draw (l2) to (l4);
	\draw (l2) to (l5);
	\draw (l3) to (l4);
	\draw (l3) to (l5);
	\draw (l4) to (l5);
	
	\node (r1) at (1.5,2.5) {};
	\node (r2) at (1.5,1.5) {};
	
	\draw (r1) to (r2);
	\draw (w) to (r1);
	\draw (w) to (r2);
	\draw (x) to (r1);
	\draw (x) to (r2);
	\draw (y) to (r1);
	\draw (y) to (r2);

\end{tikzpicture}\\[15pt]
    \begin{tikzpicture}[scale=1]
\tikzset{every node/.style={circle,
				inner sep=0pt,
				text width=1.5mm,
				align=center,
				draw=black,
				transform shape,
				fill=gray}}
		
    \node (z) at (0,0.5) {}; 
	\node (y) at (0,1) {}; 
	\draw (y) to (z);
	\node (x) at (0,3) {}; 
	\node (w) at (0,2) {}; 
	\draw (y) to (w);
	\draw (w) to (x);
	\draw (y) to [bend right=35] (x);
	
	\node (l1) at (-2,2.5) {};
	\node (l2) at (-2,1.5) {};
	\node (l3) at (-1,3) {};
	\node (l4) at (-1.25,2) {};
    \node (l5) at (-1,1) {};
    
    \draw (y) to (l3);
    \draw (y) to (l4);
    \draw (y) to (l5);
    \draw (x) to (l3);
    \draw (x) to (l4);
    \draw (x) to (l5);
    
    \draw (w) to [] (l1);
    \draw (w) to [] (l2);
    \draw (w) to [] (l3);
    \draw (w) to [] (l4);
    \draw (w) to [] (l5);
    
	\draw (l1) to (l3);
	\draw (l1) to (l4);
	\draw (l1) to (l5);
	\draw (l2) to (l3);
	\draw (l2) to (l4);
	\draw (l2) to (l5);
	\draw (l3) to (l4);
	\draw (l3) to (l5);
	\draw (l4) to (l5);
	
	\node (r1) at (1.5,2.5) {};
	\node (r2) at (1.5,1.5) {};
	
	\draw (r1) to (r2);
	\draw (w) to (r1);
	\draw (w) to (r2);
	\draw (x) to (r1);
	\draw (y) to (r2);

\end{tikzpicture} &
    \begin{tikzpicture}[scale=1]
\tikzset{every node/.style={circle,
				inner sep=0pt,
				text width=1.5mm,
				align=center,
				draw=black,
				transform shape,
				fill=gray}}
		
    \node (z) at (0,0.5) {}; 
	\node (y) at (0,1) {}; 
	\draw (y) to (z);
	\node (x) at (0,3) {}; 
	\node (w) at (0,2) {}; 
	\draw (y) to (w);
	\draw (w) to (x);
	\draw (y) to [bend right=35] (x);
	
	\node (l1) at (-2,2.5) {};
	\node (l2) at (-2,1.5) {};
	\node (l3) at (-1,3) {};
	\node (l4) at (-1.25,2) {};
    \node (l5) at (-1,1) {};
    
    \draw (y) to (l4);
    \draw (y) to (l5);
    \draw (y) to [bend left=45] (l2);
    \draw (x) to (l3);

    \draw (w) to [] (l1);
    \draw (w) to [] (l2);
    \draw (w) to [] (l3);
    \draw (w) to [] (l4);
    \draw (w) to [] (l5);
    
	\draw (l1) to (l2);
	\draw (l1) to (l3);
	\draw (l1) to (l4);
	\draw (l1) to (l5);
	\draw (l2) to (l3);
	\draw (l2) to (l4);
	\draw (l2) to (l5);
	\draw (l3) to (l4);
	\draw (l3) to (l5);
	\draw (l4) to (l5);
	
	\node (r1) at (1.5,2.5) {};
	\node (r2) at (1.5,1.5) {};
	
	\draw (r1) to (r2);
	\draw (w) to (r1);
	\draw (w) to (r2);
	\draw (y) to (r2);

\end{tikzpicture}\\
    \end{tabular}
    \caption{Four graphs with $11$ vertices which do not have a $2$-PDS partition}
    \label{figure: 11 vertices bad guys}
\end{figure}

In \cref{figure: 11 vertices bad guys}, we present  four graphs with $11$ vertices without a $2$-PDS partition.
These graphs have an odd  number of vertices, hence they do not belong to $\mathcal{G}$.
To prove that they do not have a $2$-PDS partition, one can notice that, like the graphs in $\mathcal{G}$, they have a pendant vertex $z$ connected to a vertex $y$, and a vertex $w$ connected to all the vertices except the pendant vertex.
As a result, the vertex $z$ is satisfied if and only if it belongs to the same PDS as $y$, and thus $w$ must be in the other PDS. 
The rest of the proof can be done by case distinction.

\subsection{Disconnected $2$-PDS partition}\label{subsection: disconnected 2-PDS partition}

Now, we present an infinite family of graphs where each graph admits a disconnected $2$-PDS partition, but not a connected one.
The existence of such graphs was left as an open problem in \cite{BCP17}.

\begin{definition}
We define the infinite class of graphs $\mathcal{H}$ such that, if $G=(V,E) \in \mathcal{H}$, then
\begin{compactitem}
    \item $V := W \cup \{\alpha_1,\beta_1,\alpha_2,\beta_2\}$, where $W$ is a clique of odd size at least $7$;
    \item $\exists{} x,y \in W$ such that $\{x,\alpha_1\}, \{x,\beta_1\}, \{x,\beta_2\}, \{y,\alpha_2\}, \{y,\beta_2\}, \{y,\beta_1\} \in E$;
    \item $\{\alpha_1,\beta_1\}, \{\alpha_2,\beta_2\} \in E$.
\end{compactitem}
\end{definition}

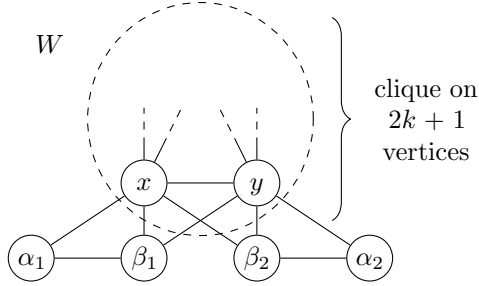
\begin{figure}[htbp]
    \centering
    \begin{tikzpicture}[scale=1.0]
		\tikzset{every node/.style={circle,
				inner sep=0pt,
				text width=5mm,
				align=center,
				draw=black,
				transform shape,
				fill=white}
		}
	
	
	
	\node (b1) at (-2.25,-1) {$\mathstrut{\alpha_1}$};
	\node (b2) at (-0.75,-1) {$\mathstrut{\beta_1}$};
	\draw (b1) to (b2);
	\node (b3) at (0.75,-1) {$\mathstrut{\beta_2}$};
	\node (b4) at (2.25,-1) {$\mathstrut{\alpha_2}$};
    \draw (b3) to (b4);
    
	\node (c1) at (-.75,0) {$\mathstrut{x}$};
	\node (c2) at (0.75,0) {$\mathstrut{y}$};

	\draw (c1) to (-.75,.5);
    \draw[dashed] (-.75,.5) to ++(0,.5);
    \draw (c1) to (-.5,.5);
    \draw[dashed] (-.5,.5) to ++(0.25,.5);
    
    \draw (c2) to (.75,.5);
    \draw[dashed] (.75,.5) to ++(0,.5);
    \draw (c2) to (.5,.5);
    \draw[dashed] (.5,.5) to ++(-0.25,.5);

	\draw (c1) to (b1);
	\draw (c2) to (b4);
	\draw (c1) to (b2);
	\draw (c1) to (b3);
	\draw (c2) to (b2);
	\draw (c2) to (b3);
	
	\draw (c1) to (c2);
	

	\draw[dashed] (0,.85) ellipse (1.5cm and 1.55cm);
	\draw [decorate,decoration={brace,amplitude=7pt}] (1.75,2.2) -- (1.75,-.50);
	\draw node[draw=none,rectangle, text width=2cm, text centered] at (3,0.85) {clique on $2k+1$ vertices};
	\draw node[draw=none,rectangle, text centered] at (-2,1.85) {$W$};
\end{tikzpicture}
    \caption{A schematic representation of a graph in $\mathcal{H}$.}
    \label{figure: 11 vertices disconnected}
\end{figure}

Compared to the graphs in $\mathcal{G}$, each graph in $\mathcal{H}$ has an odd number of vertices, the smallest one has $11$ vertices.

\begin{theorem}
    All graphs in $\mathcal{H}$ do not have a connected $2$-PDS partition, but have a disconnected one.
\end{theorem}
\begin{proof}
    Let $G=(V,E) \in \mathcal{H}$.
    Suppose that $G$ has a connected $2$-PDS partition $\{A,B\}$.
    If $A \subseteq W$, then we have two cases: either $A = W$ but then $G[B]$ is disconnected, or $A \subset W$ but then a vertex in $W \setminus A$ is not satisfied in $B$.
    Hence, $A \nsubseteq W$ and similarly $B \nsubseteq W$.
    Consequently, to guarantee the connectivity of $G[A]$ and $G[B]$, the vertices $x$ and $y$ must be in different parts of the partition.
    Therefore, we assume without loss of generality that $x \in A$ and $y \in B$.
    
    If $\alpha_1 \in B$, then $y$ is not satisfied since it is connected to each vertex in $A$.
    Similarly, $\alpha_2$ cannot belong to $A$ since otherwise $x$ is not satisfied.
    As a result, we only have to consider the possible cases for $\beta_1$ and $\beta_2$, knowing that $x, \alpha_1 \in A$ and $y, \alpha_2 \in B$. 
    
    If $\beta_1 \in A$ and $\beta_2 \in B$, then consider two vertices $a \in (W \setminus \{x\}) \cap A$ and $b \in (W \setminus \{y\}) \cap B$.
    The vertex $a$ is satisfied in $A$ if and only if
    \[
        \frac{d_A(a)}{|A|-1} = \frac{|A|-3}{|A|-1} \geq \frac{|B|-2}{|B|} = \frac{d_B(a)}{|B|}\,,
    \]
    which implies that $|A| \geq |B|+1$.
    Similarly, the vertex $b$ is satisfied in $B$ if and only if $|A| \leq |B|-1$, which is a contradiction.
    
    If $\beta_1, \beta_2 \in A$, then the vertex $\beta_2$ is satisfied in $A$ if and only if
    \[
        \frac{d_A(\beta_2)}{|A|-1} = \frac{1}{|A|-1} \geq \frac{2}{|B|} = \frac{d_B(\beta_2)}{|B|}\,,
    \]
    which implies that $|A| \leq \frac{|B|}{2}+1$.
    Moreover, the vertex $\alpha_2$ is satisfied in $B$ if and only if
    \[
        \frac{d_B(\alpha_2)}{|B|-1} = \frac{1}{|B|-1} \geq \frac{1}{|A|} = \frac{d_A(\alpha_2)}{|A|}\,,
    \]
    which implies that $|A| \geq |B|-1$.
    We then obtain $|B|-1 \leq |A| \leq \frac{|B|}{2}+1$, and therefore $|B| \leq 4$.
    Thus, $|A| \leq 3$, which is not possible since $|V| \geq 11$.
    Similar arguments can be used to prove that $\beta_1$ and $\beta_2$ cannot both belong to $B$.
    
    We conclude that $G$ does not have a connected $2$-PDS partition.
    However, it is easy to see that, if $A := \{\alpha_1,\beta_1,\alpha_2,\beta_2\}$ and $B := V \setminus A$, then $\{A,B\}$ is a disconnected $2$-PDS partition of $G$.
\end{proof}

\section{Conclusion and further work}\label{section: conclusion}

The definition of a proportionally dense subgraph is based on a combination of local and global properties, where each vertex has to satisfy a condition depending not only on its degree but also on the size of the subgraph.
This property makes the problem  complex from an algorithmic point of view and requires a novel approach.

Our infinite families of graphs bring a new insight into the existence of $2$-PDS partitions in graphs, with and without constraint of connectivity.
Further research may investigate the structural characterisations of graphs with or without a (connected) $2$-PDS partition.
These results can help to answer the following important question: what is the complexity of deciding whether a graph admits a (connected) $2$-PDS partition?
\biboptions{sort&compress,numbers}
\bibliographystyle{abbrvnat}
\bibliography{refs.bib}

\end{document}